%% file: revised_draft_R1.tex
\documentclass[final]{IEEEtran}
\IEEEoverridecommandlockouts

\usepackage{amsthm,amssymb,graphicx,multirow,amsmath,color,amsfonts}%,ulem}
\usepackage[update,prepend]{epstopdf}
\usepackage[noadjust]{cite}
\usepackage[latin1]{inputenc}
\usepackage{tikz}
\usetikzlibrary{arrows,calc}		% Optional ticks libraries
\usepackage{bbm} % for \mathbbm{1}
\usepackage{pdfpages}
\usepackage{tabulary}
\usepackage{multirow}
\usepackage{comment}
\usepackage{subcaption} % For subfigures
\usepackage{caption}  % For customizing captions
\usepackage[ruled,vlined,linesnumbered]{algorithm2e}
\pdfoutput=1
\usepackage{svg}
\usepackage{enumitem,kantlipsum}
\usepackage{graphicx} % Required for resizing
\usepackage{balance}
\def\chb#1{{\color{blue}#1}}

\renewcommand{\chb}[1]{\textcolor{black}{#1}}

\include{notation}

\allowdisplaybreaks % Allows breaking of eqnarray over multiple pages (avoids unnecessary blanks in the document before eqnarray)

\begin{document}
\graphicspath{{./figures/}}
\title{%\huge 
% \fontsize{23}{19}\selectfont LEO-based Joint Delay and Carrier-Phase Positioning: Design Insights and Convergence Analysis}
% \fontsize{23}{19}\selectfont LEO-based Joint Delay and Carrier-Phase Positioning for 6G: Design Insights and {Comparison with GNSS}}
\huge LEO-based Carrier-Phase Positioning for 6G: Design Insights and {Comparison with GNSS}}
\author{
Harish K. Dureppagari, Harikumar Krishnamurthy, Chiranjib Saha, Xiaofeng Wang, Alberto Rico-Alvari\~{n}o, R. Michael Buehrer, Harpreet S. Dhillon
\thanks{H. K. Dureppagari, R. M. Buehrer, and H. S. Dhillon are with Wireless@VT, Department of ECE, Virginia Tech, Blacksburg, VA 24061, USA. Email: \{harishkumard, rbuehrer, hdhillon\}@vt.edu. H. Krishnamurthy, C. Saha, X. Wang, and A. Rico Alvari\~{n}o are with the Qualcomm Standards and Industry Organization, Qualcomm Technologies Inc., San Diego, CA 92121, USA. Email: \{harkris, csaha, wangxiao, albertor\}@qti.qualcomm.com. The support of the US NSF (Grant CNS-2107276) is gratefully acknowledged.
}
\vspace{-1mm}
}

\maketitle

\begin{abstract}
The integration of non-terrestrial networks (NTN) into 5G new radio (NR) enables a new class of positioning capabilities based on cellular signals transmitted by Low-Earth Orbit (LEO) satellites. % Compared to Global Navigation Satellite Systems (GNSS), LEO-based positioning benefits from stronger link budgets, wider usable bandwidths, and dense constellations with faster satellite motion. 
In this paper, we investigate joint delay-and-carrier-phase positioning for LEO-based NR-NTN systems and provide a convergence-centric comparison with Global Navigation Satellite Systems (GNSS). We show that the rapid orbital motion of LEO satellites induces strong temporal and geometric diversity across observation epochs, thereby improving the conditioning of multi-epoch carrier-phase models and enabling significantly faster integer-ambiguity convergence. To enable robust carrier-phase tracking under intermittent positioning reference signal (PRS) transmissions, we propose a dual-waveform design that combines wideband PRS for delay estimation with a continuous narrowband carrier for phase tracking. Using a realistic simulation framework incorporating LEO orbit dynamics, we demonstrate that LEO-based joint delay-and-carrier-phase positioning achieves cm-level accuracy with convergence times on the order of a few seconds, whereas GNSS remains limited to meter-level accuracy over comparable short observation windows. These results establish LEO-based cellular positioning as a strong complement and potential alternative to GNSS for high-accuracy positioning, navigation, and timing (PNT) services in future wireless networks.
\end{abstract}

\vspace{-1mm}
\section{Introduction}\label{sec:intro}
Positioning has been an integral capability of wireless networks, initially driven by regulatory requirements and later by the growing demand for location-aware services. The introduction of NTN into 5G NR, starting from Release 17, enables cellular services via satellites in low-earth orbit (LEO), medium-earth orbit (MEO), and geostationary-earth orbit (GEO)~\cite{dureppagari_ntn_10355106,dureppagari_leo_11049853}. While this development has allowed emerging LEO constellations, such as Starlink and OneWeb, to provide global enhanced mobile broadband (eMBB) and Internet of Things (IoT) services, it also creates an opportunity to enable satellite-based positioning using cellular signals. LEO satellites, typically orbiting at altitudes of around 600 km, offer substantially stronger link budgets and higher received signal strength than GNSS satellites operating at around 20,200 km. This improved signal strength, combined with access to wider bandwidths, can significantly reduce time-to-first-fix (TTFF) relative to GNSS. Moreover, LEO-based cellular positioning has the potential to reduce reliance on dedicated GNSS radios in user equipment (UE), which, as of Release 19, are still required for NR-NTN connectivity, thereby reducing power consumption and hardware complexity. 

%With GNSS remaining a cornerstone of PNT services, a fundamental question remains: can LEO constellations serve as a viable alternative or complement to GNSS for PNT using 6G cellular infrastructure? 
In our prior work~\cite{dureppagari_ntn_10355106}, we presented our vision for NTN-based localization, outlining prospective study cases for accurate positioning in 6G, highlighting the role of LEOs, and identifying key challenges in integrating such systems into 5G and beyond. Building on this vision, in our subsequent work~\cite{dureppagari_leo_11049853}, we investigated the feasibility of repurposing communication-centric LEO constellations for positioning, presented NR-NTN-compliant design guidelines, and demonstrated that sub-10m accuracy can be achieved using delay-based methods with 10 MHz bandwidth, yielding performance comparable to GNSS~\cite{Edkgps2005}. Traditional delay-based positioning methods cannot achieve cm-level accuracy due to several limitations, including atmospheric biases, multipath, satellite clock errors, and UE clock errors. %These biases need to be estimated and corrected to achieve high positioning accuracy. 
Moreover, positioning using delay-based methods is limited by bandwidth and receiver sampling rate. This motivates the need for more advanced positioning techniques that exploit carrier-phase measurements, which offer orders-of-magnitude higher ranging precision and form the basis of high-accuracy satellite positioning systems.

{\em Prior Art.} High-accuracy satellite positioning has traditionally relied on carrier-phase measurements~\cite{teunissen2017springer}, where cm-level accuracy is achieved through phase continuity and integer ambiguity resolution, typically aided by code-phase measurements. Integer ambiguity is an unknown integer number of full carrier cycles accumulated along the signal propagation path, which cannot be directly observed from instantaneous carrier-phase measurements. Joint processing of carrier-phase measurements across multiple observation epochs helps resolve this ambiguity, while code-phase measurements provide coarse range information that significantly accelerates ambiguity resolution. With the integration of NTN into NR, this paradigm has recently attracted interest in the context of LEO-based positioning. Studies in~\cite{starlink_first_carrier_phase,unveiling_starlink_pnt} have shown that carrier-phase tracking is feasible using communication-centric LEO signals-of-opportunity (SOP), such as Starlink, by coherently tracking beat carrier phase and recovering phase-based measurements suitable for positioning from waveforms not originally designed for PNT. Subsequent studies have further extended this line of work to study differential carrier-phase positioning and ambiguity resolution in LEO megaconstellations, as well as theoretical formulations of ambiguity fixing under carrier-phase models that account for frequency variations caused by high Doppler dynamics~\cite{leo_differential_carrier_phase,ambiguity_resolved_leo_jog}%,ambiguity_fixing_freq_varying}.

However, existing studies primarily focus on SOP-based architectures, assume long observation windows (on the order of several minutes), do not discuss convergence behavior or systematic comparisons with GNSS, and, importantly, do not address NR-NTN constraints. Although carrier-phase positioning has also been explored in terrestrial 5G-Advanced NR~\cite{10475845}, these techniques cannot be directly extended to NR-NTN due to fundamentally different dynamics, including high Doppler and Doppler-rate effects and communication-centric constellation designs. Consequently, NR-NTN literature has largely focused on delay-based positioning, which alone cannot achieve carrier-phase-grade accuracy. To this end, our work lays the foundation for using a joint delay-and-carrier-phase positioning framework tailored to LEO-based NR-NTN systems, demonstrating rapid convergence and cm-level accuracy by explicitly leveraging the fast satellite motion inherent to LEO constellations and positioning LEO-based PNT as a potential alternative to GNSS.

{\em Contributions.} In this paper, we investigate the joint exploitation of carrier-phase and delay-based measurements for high-precision positioning, aiming to achieve cm-level accuracy. We begin by describing how the carrier phase is measured and tracked, and analyze the impact of satellite motion on carrier-phase evolution and the accuracy of Doppler-based phase approximations in both LEO and GNSS systems. We then study the temporal evolution of the phase transformation matrix obtained by jointly processing carrier-phase measurements over multiple epochs, with particular emphasis on its conditioning properties quantified by the condition number, and compare its convergence behavior for LEO and GNSS constellations. Next, we present a comprehensive comparative analysis of integer ambiguity resolution and positioning performance under delay-only positioning and joint delay-and-carrier-phase positioning frameworks for both LEO- and GNSS-based systems. Through our analysis, we demonstrate that the rapid satellite motion in LEOs significantly enhances geometric diversity across observation epochs, leading to faster ambiguity resolution and substantially reduced convergence time compared to GNSS. Subsequently, we show that LEO-based joint delay-and-carrier-phase positioning achieves cm-level accuracy with convergence times on the order of a few seconds, thereby delivering superior performance compared to GNSS within short observation windows. We conclude our discussion by establishing that LEO-based positioning is a potential alternative to GNSS, serving as a standalone, high-accuracy PNT solution in future wireless networks.

\section{Carrier-Phase Measurements \& Tracking}\label{sec:carrier_phase_tracking}
\begin{figure}[t]
    \centering
    \includegraphics[width=0.80\linewidth]{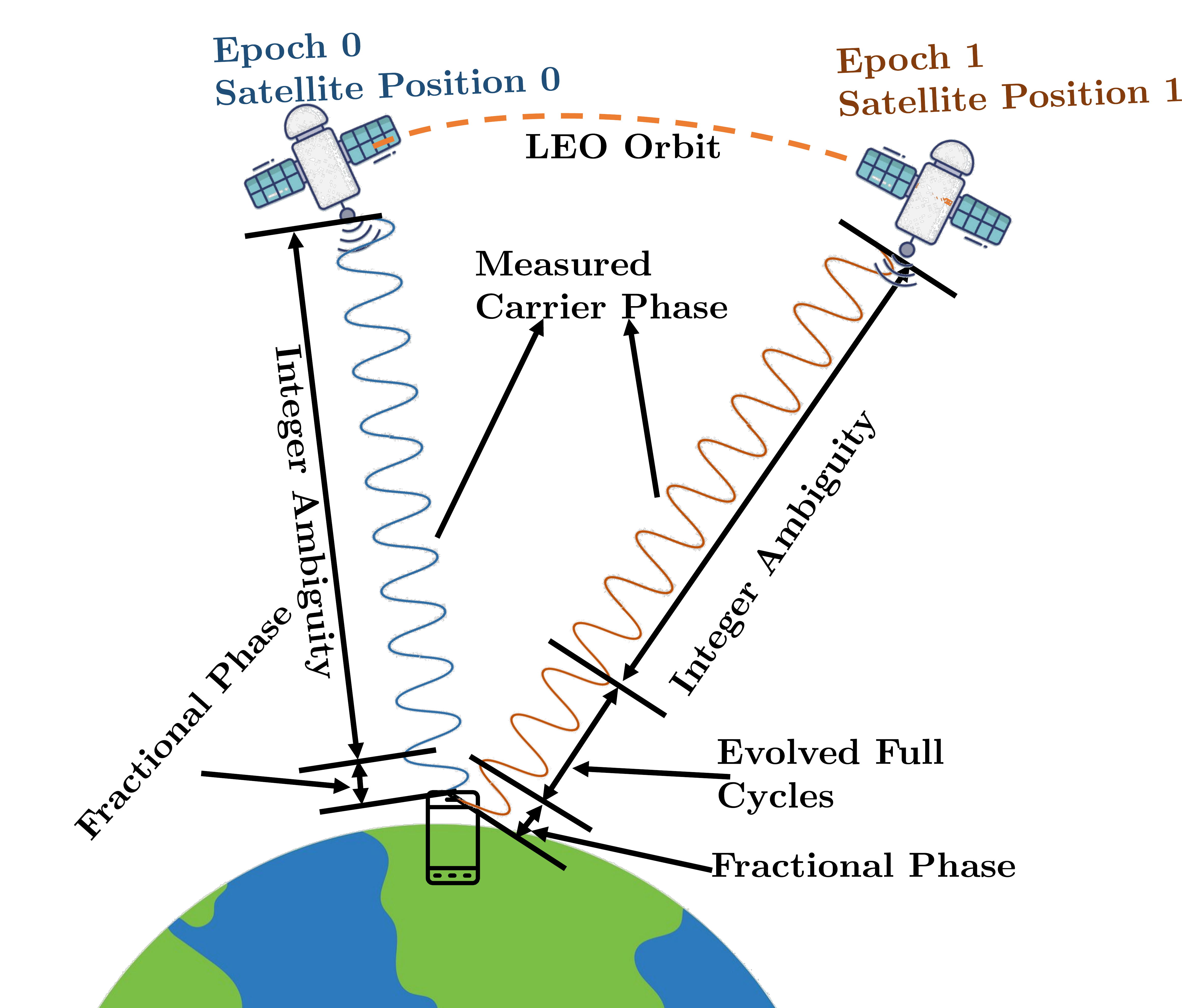}
    \captionsetup{font=small}
    \caption{{\em Introduction to Carrier Phase}: Carrier phase measurements over time.}
    \label{fig:carrier_phase_intro}
\vspace{-18pt}
\end{figure}

Carrier-phase measurements enable range estimation with a precision on the order of the carrier wavelength, offering several orders of magnitude improvement over delay-based measurements. Conceptually, the carrier phase represents the accumulated phase of the received sinusoidal carrier relative to the transmitted signal. As illustrated in Fig~\ref{fig:carrier_phase_intro}, the carrier phase is measured over multiple epochs as the satellite traverses in its orbit. At any given observation epoch, the measured carrier phase consists of two key components: a fractional phase that can be directly observed, and an unknown integer number of full carrier cycles that have elapsed along the signal propagation path. This unknown integer offset, commonly referred to as the integer ambiguity, is constant over time as long as phase continuity is preserved. At an initial observation epoch, only the fractional part of the carrier phase is measurable, while the integer ambiguity remains unknown. As the satellite moves along its orbit and subsequent observations are made, the total carrier phase accumulates full carrier cycles evolved and a new fractional phase at each epoch. By jointly processing carrier-phase measurements across multiple epochs, it becomes possible to resolve the integer ambiguity, i.e., achieve highly accurate range estimation, which in turn enables cm-level positioning accuracy. In practice, delay- or code-phase measurements, even when relatively coarse compared to carrier-phase measurements, play a critical role in accelerating integer ambiguity resolution. By providing an initial range estimate, delay-based measurements significantly reduce the feasible integer ambiguity search space, enabling much faster and more reliable ambiguity resolution than would be possible using carrier-phase measurements alone. We discuss the complementary use of delay and carrier-phase measurements in detail in the next section.

However, carrier-phase tracking poses several fundamental challenges, particularly in LEO-NTN systems. First, maintaining phase continuity without cycle slips is difficult due to noise, oscillator instability, multipath, and low signal-to-noise ratio (SNR), which can cause loss of tracking lock, reset the integer ambiguity, and invalidate the accumulated phase history, requiring reinitialization and degrading positioning performance. Second, accurate frequency estimation is critical, as residual frequency errors cause rapid phase drift and large phase uncertainties within tens of ms in high-mobility LEO scenarios, making continuous tracking necessary between intermittent measurements. For instance, with a 40~ms PRS periodicity, the frequency error between consecutive PRS occasions, including Doppler variation, should remain below 25~Hz to avoid losing carrier-phase cycles. Third, cm-level positioning requires reliable multi-epoch phase continuity for integer ambiguity resolution. Inaccurate phase evolution across epochs rapidly expands the ambiguity search space, potentially making ambiguity resolution infeasible, particularly for LEOs, where rapid satellite motion induces large Doppler shifts and Doppler rates and causes the carrier phase to evolve by thousands of cycles over short intervals.

\begin{figure}[t]
    \centering
    \includegraphics[width=0.8\linewidth]{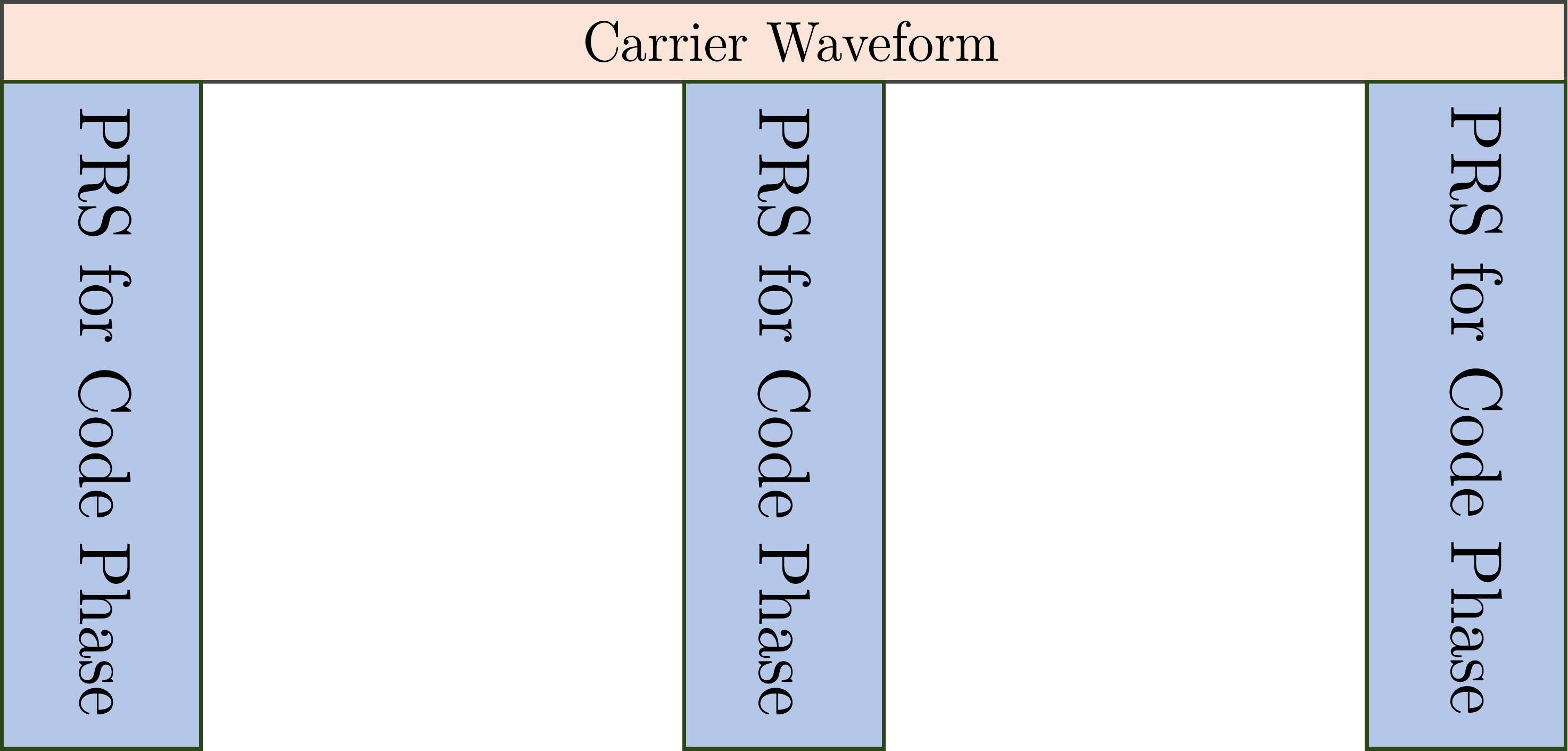}
    \captionsetup{font=small}
     \caption{Dual waveform.}
     \label{fig:dual_waveform}
\vspace{-18pt}
\end{figure}

These challenges become particularly severe in NR-NTN, where PRS is typically transmitted intermittently with configurable periodicity and muting patterns to limit overhead and enable multiplexing with communication services. Wideband PRS transmissions are typically leveraged to obtain high-resolution code-phase measurements. While wideband PRS is well-suited for estimating signal delay and providing absolute timing references, its intermittent nature prevents continuous observation of carrier-phase evolution. As a result, the phase change between consecutive PRS occasions becomes unobservable, leading to severe integer ambiguity growth and a high likelihood of cycle slips. To address these limitations in LEO-based NR-NTN systems, we propose a dual-waveform design that combines wideband PRS with a continuous narrowband carrier waveform as illustrated in Fig.~\ref{fig:dual_waveform}. While PRS transmissions are intermittent wideband bursts, the narrowband carrier waveform is continuous and spans the entire interval between consecutive PRS occasions. %The fundamental motivation behind this design is to decouple the roles of code-phase (delay) estimation and carrier-phase tracking, which have fundamentally different waveform and transmission requirements. %This continuous carrier enables uninterrupted carrier-phase and Doppler tracking across epochs, thereby allowing reliable phase propagation through PRS gaps. 

\chb{It is important to emphasize that the novelty of the proposed dual-waveform architecture lies in adapting the NR positioning framework to support high-accuracy carrier-phase positioning under the rapid phase dynamics of LEO satellites. Unlike conventional NR positioning, which primarily relies on intermittent wideband PRS~\cite{10475845}, the proposed architecture decouples two fundamentally different requirements: high-resolution code-phase (delay) estimation and continuous carrier-phase tracking. The former benefits from large bandwidth but does not require continuous transmission, whereas the latter requires temporal continuity to reliably track carrier phase, Doppler, and Doppler rate but needs only a small bandwidth. Accordingly, the proposed design retains intermittent wideband PRS for ranging while introducing a continuous narrowband waveform for phase and frequency tracking between PRS occasions. This separation avoids continuous transmission of a resource-intensive wideband PRS and enables continuous carrier-phase positioning in communication-centric NR-NTN deployments with limited additional spectrum and payload resources. From a practical deployment perspective, the continuous phase-tracking waveform can be supported through a dedicated phase-stable RF chain or integrated with the existing satellite transmit chain.}

\chb{A representative continuous narrowband waveform can occupy as little as one physical resource block (PRB) (180~kHz at 15~kHz SCS). Depending on spectrum availability, it can be placed at the carrier edge or within suitable guard-band resources, minimizing resources otherwise allocated for communication. Alternatively, one PRB within the active NR-NTN carrier can be reserved for continuous phase tracking, resulting in a small communication-resource overhead (e.g., less than 1\% of the nominal bandwidth for a 20~MHz carrier). For multi-satellite positioning, phase-tracking waveforms from multiple visible satellites can share the same narrow subband using satellite-specific sequences and reduced transmit power, confining the additional positioning bandwidth to one PRB and allowing the UE to continuously monitor a single frequency region. Alternatively, satellite-specific waveforms can be frequency-division multiplexed (FDM) when additional spectrum is available, at the expense of increased bandwidth, Doppler-dependent guard spacing, and receiver-monitoring complexity. Similar reduced-power, common-subband, and FDM-based positioning architectures have recently been investigated for integrated LEO communication and PNT systems~\cite{esa_ntn_com_pnt_p1}. The continuous-waveform transmit power can be configured based on the SNR required for reliable phase tracking while limiting interference on communication signals. Detailed optimization of PRB and power allocation is beyond the scope of this work.}

In the context of a comparative analysis of GNSS- and LEO-based carrier-phase positioning, we first investigate how satellite motion impacts carrier-phase evolution and assess the accuracy of Doppler-based phase approximation relative to the true carrier phase. A Doppler-based approximation typically models the carrier phase as a linear function of the estimated Doppler, whereas the true carrier phase evolution is generally nonlinear due to time-varying Doppler, particularly for LEO satellites. For this analysis, we consider a representative LEO satellite orbiting at 600 km and a GNSS satellite at around 20,200 km, both initially positioned at nadir, corresponding to a 90\(^\circ\) elevation angle at the UE. LEO satellites orbit at velocities on the order of 7.5 km/s, while GNSS satellites orbit at approximately 3.9 km/s. The higher orbital velocity of LEO satellites results in significantly larger Doppler shifts and, more importantly, substantially higher Doppler rates as the satellite traverses its orbit. In particular, at 90\(^\circ\) elevation angle, the Doppler variation reaches its maximum: the Doppler rate is approximately  600 Hz/s for the LEO satellite at a 2 GHz carrier frequency, whereas it is approximately 3.9 Hz/s for the GNSS satellite operating at 1575.42 MHz (GPS L1).

\begin{figure}[t]
     \centering
     \begin{subfigure}[b]{0.450\columnwidth}
        \centering
        \includegraphics[width=\textwidth]{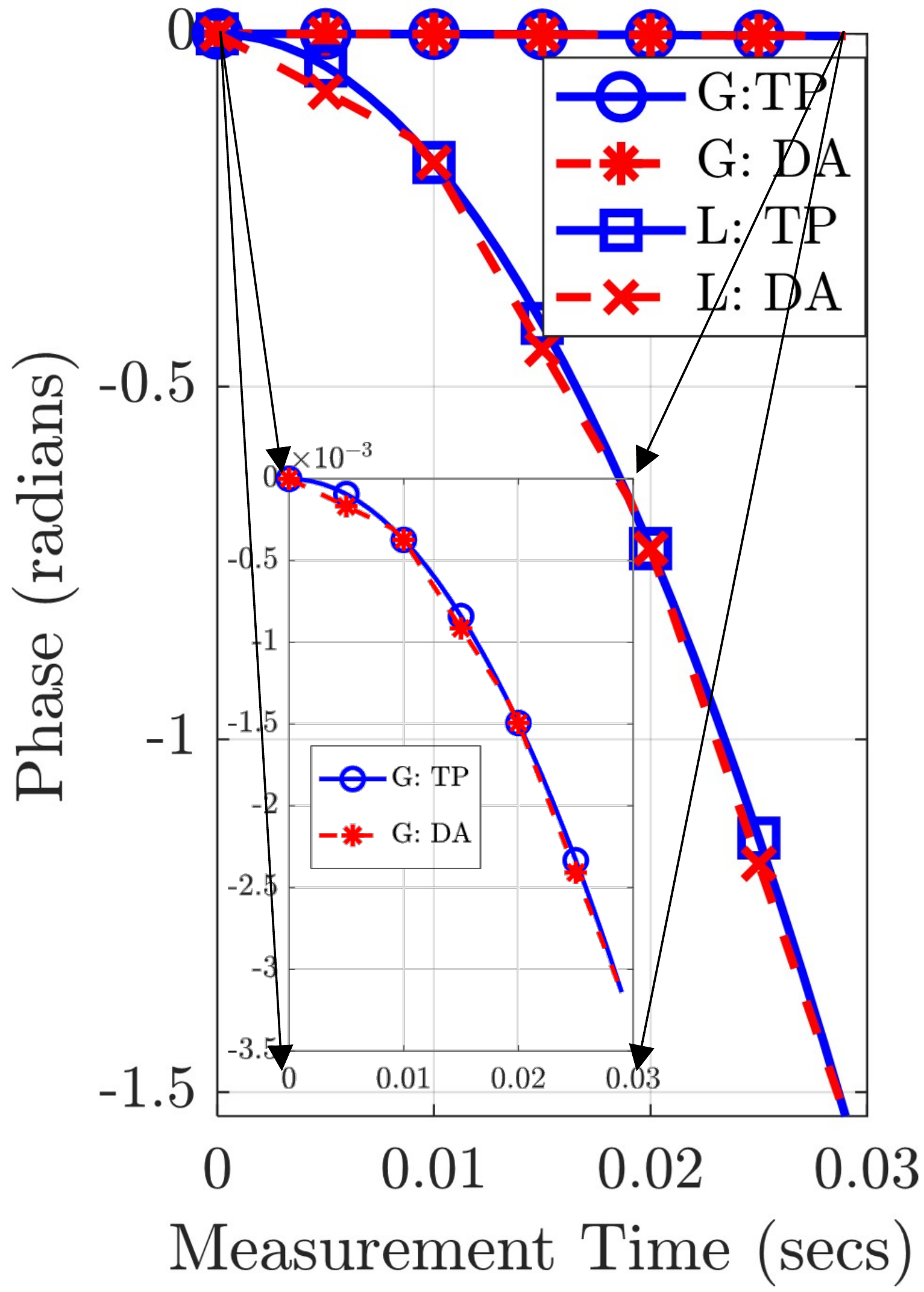}
        \captionsetup{font=small}
         \caption{Phase evolution.}
         \label{fig:phase_evol}
     \end{subfigure}
     \hfill
     \begin{subfigure}[b]{0.520\columnwidth}
        \centering
        \includegraphics[clip, trim=2.81cm 0.1cm 3.25cm 0.60cm, width=\textwidth]{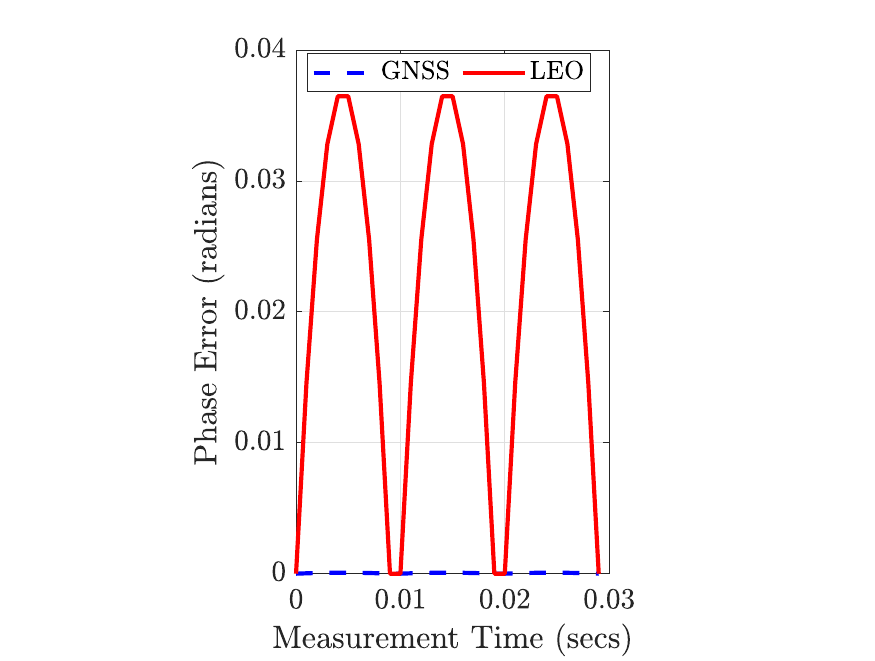}
        \captionsetup{font=small}
         \caption{Phase error.}
         \label{fig:phase_err}
     \end{subfigure}
     \captionsetup{font=small}
     \caption{{\em Carrier Phase evolution and error: GNSS vs. LEO:} a) True phase and Doppler approximation over time (G - GNSS, L - LEO, TP - True Phase, and DA - Doppler Approximation); b) error (difference between TP and DA) due to Doppler approximation over time.}
     \label{fig:phase_evol_err}
\vspace{-18pt} \end{figure}

Fig.~\ref{fig:phase_evol} presents the temporal evolution of the true carrier phase and its Doppler-based approximation for GNSS and LEO satellites over a 30~ms observation window, with Doppler estimates updated every 10~ms. As anticipated, the carrier phase evolves much faster for LEOs due to their significantly higher orbital velocity. Although both systems exhibit similar overall trends, the deviation between the true and Doppler-approximated phases is noticeably larger for LEOs, indicating that the linear Doppler approximation is more accurate for GNSS due to its slower Doppler variation. Fig.~\ref{fig:phase_err} further quantifies this deviation as the phase approximation error, which is substantially higher for LEOs due to their larger Doppler variations. Importantly, while rapid LEO motion increases this approximation error, it also provides greater temporal diversity and faster carrier-phase convergence, thereby enabling higher positioning accuracy than GNSS, which we discuss in detail in Section~\ref{sec:convergence_positioning}. Even with linear models, the range of phase approximation errors decreases as the satellite moves away from nadir and the Doppler variation decreases. Furthermore, LEO phase approximation errors can be reduced using nonlinear phase models incorporating Doppler rate and higher-order motion terms.

\section{Convergence Analysis and Positioning Performance}\label{sec:convergence_positioning}
In this section, we present a comprehensive convergence and positioning performance comparison of GNSS- and LEO-based positioning. The convergence analysis is conducted in two stages to highlight the fundamental differences between the systems. First, we examine the conditioning of the phase-evolution transformation matrix when carrier-phase measurements are jointly processed across multiple epochs, thereby assessing the impact of satellite motion and geometric diversity. Second, we analyze integer ambiguity convergence using joint delay-and-carrier-phase measurements, emphasizing the impact of satellite dynamics. We then evaluate positioning performance under delay-only and joint delay-and-carrier-phase frameworks, demonstrating the gains from incorporating carrier-phase measurements and how integer ambiguity resolution enables higher positioning accuracy and faster convergence for LEOs than GNSS. For evaluation, we assume repurposing communication-focused LEO constellations for positioning, consistent with the NR-NTN context considered in this work. Adapting such constellations for high-accuracy positioning involves several challenges that are beyond the scope of this paper; interested readers are referred to~\cite{dureppagari_leo_11049853} for a comprehensive discussion.

Carrier-phase positioning can be implemented using several well-established frameworks, including conventional static, semi-kinematic, and real-time kinematic (RTK) positioning~\cite{teunissen2017springer}. In this work, we adopt conventional static positioning for evaluation, assuming a stationary reference receiver with a perfectly known location and a reference satellite, typically the one with the strongest SNR, similar to delay-based time-difference-of-arrival (TDOA) positioning. Carrier-phase measurements are double-differenced across satellites and receivers with respect to the reference satellite and receiver. Differencing with respect to the reference satellite suppresses UE clock drift, while differencing with respect to the reference receiver eliminates satellite clock drift and reduces atmospheric errors. The unknown parameter vector consists of the UE coordinates relative to the reference receiver and the corresponding carrier-phase integer ambiguities. Although we adopt static positioning for evaluation, the choice of framework does not affect the key insights and conclusions of this work.

Carrier-phase integer ambiguities typically do not have a time index and are constant throughout the observation window as long as phase continuity is maintained and no cycle slips occur. A cycle slip causes an abrupt jump in the ambiguity by an integer multiple of the carrier wavelength and can be reliably detected using standard hypothesis-testing techniques~\cite{teunissen2017springer,teunissen1997gps}. Upon detection, cycle slips can be handled either by adapting the ambiguity parameters to preserve temporal consistency or by discarding the carrier-phase measurements collected prior to the cycle slip and reinitializing the measurement process. \chb{More advanced cycle-slip mitigation can be incorporated through Kalman-filter-based tracking, where carrier-phase, Doppler, and Doppler-rate states are jointly predicted, and abrupt phase deviations are used to detect loss of phase continuity. Once detected, the corresponding integer-cycle discontinuity can be corrected and the ambiguity state updated; otherwise, the affected ambiguity can be reinitialized. The proposed dual-waveform architecture naturally supports this process, with the continuous narrowband waveform enabling cycle-slip monitoring between PRS occasions and the wideband PRS providing periodic delay measurements to re-anchor and validate the tracking states.}

Note that the double-differenced carrier-phase measurements over multiple epochs can be expressed as a linear function of the unknown parameter vector through a phase transformation matrix. By jointly processing the double-differenced measurements across multiple observation epochs, both the relative position and the integer ambiguities can be estimated. Finally, an estimate of the UE location can be obtained by combining the estimated relative coordinates with the known location of the reference receiver~\cite{teunissen2017springer}.  \chb{It is also important to note that, while instantaneous satellite geometry affects positioning accuracy, conventional geometric dilution of precision (GDOP) alone is insufficient to characterize multi-epoch joint delay-and-carrier-phase positioning, as it does not capture the temporal phase diversity that determines integer ambiguity convergence. Therefore, we examine the temporal carrier-phase evolution, conditioning of the accumulated multi-epoch phase-evolution matrix, integer ambiguity convergence, and the resulting positioning performance to characterize the convergence benefits enabled by rapid LEO satellite motion.}

\subsection{Conditioning of Phase Evolution Transformation Matrix}\label{sec:condition_number}
Before analyzing the conditioning of the phase-evolution transformation matrix, we first describe the evaluation setup. For the LEO case, we consider a typical communication-focused constellation with orbits at 600 km altitude and an inclination of 70$^\circ$, comprising 30 orbits with 28 satellites per orbital plane, for a total of 840 satellites. These parameters are chosen to reflect realistic LEO constellation designs that ensure global coverage and adequate satellite visibility~\cite{9840374}. For the GNSS case, satellite positions are generated using the MATLAB built-in GNSS toolbox, which computes satellite positions and velocities for MEO constellations based on the orbital parameters specified in the IS-GPS-200N interface control document{~\cite{ISGPS200N}}. The GNSS satellites are modeled at an inclination of 55$^\circ$. In both cases, the UE is assumed to be located at latitude 0 and longitude 0. A satellite is considered visible if the elevation angle with respect to the UE exceeds 15$^\circ$. The observation epoch is assumed to be 10ms. At each observation epoch, the set of visible satellites is identified, and the corresponding carrier-phase measurements at the UE are collected to construct the phase-evolution transformation matrix for that epoch. Carrier-phase measurements are then accumulated over multiple epochs and jointly processed to form a unified phase transformation matrix that captures the temporal evolution of the carrier phase. The conditioning of the resulting phase transformation matrix is evaluated as the number of observation epochs increases. Notably, the set of visible satellites evolves across epochs, particularly in the LEO case, due to the significantly faster satellite motion and frequent changes in satellite geometry.
\begin{figure}[!tb]
    \centering
    \includegraphics[width=0.88\linewidth]{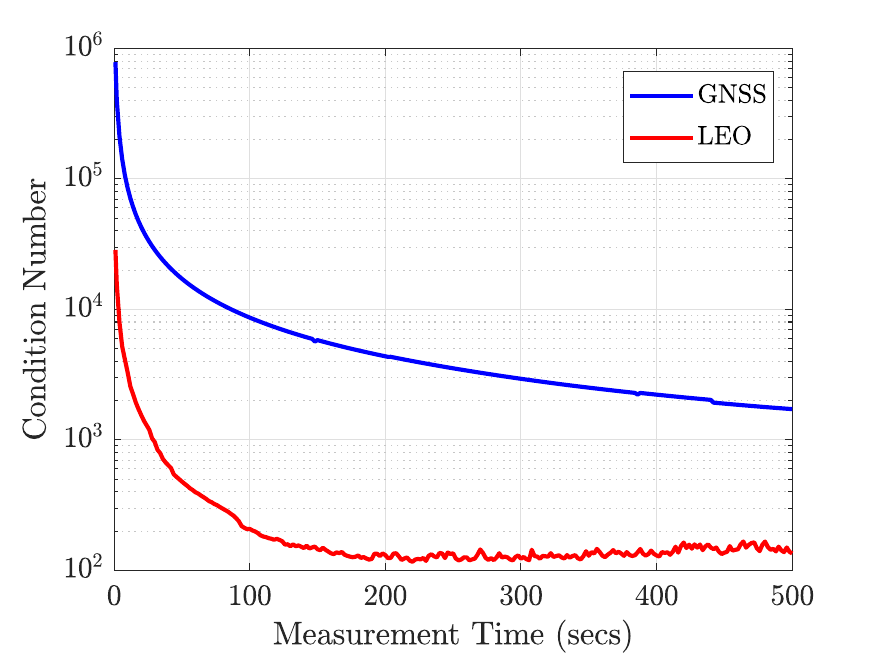}
    \captionsetup{font=small}
    \caption{{\em Condition Number Analysis: GNSS vs. LEO}: Conditioning of phase evolution transformation matrix over measurement time.}
    \label{fig:cond_analysis}
\vspace{-18pt}
\end{figure}

Fig.~\ref{fig:cond_analysis} presents the conditioning behavior of the phase-evolution transformation matrix for GNSS and LEO satellite constellations by plotting the condition number as a function of the measurement duration. The condition number of a matrix is the ratio of the largest singular value to the smallest singular value. The lower the condition number, the better the rank and conditioning. In the context of carrier-phase positioning, the condition number reflects the combined effect of UE-to-satellite geometry and the temporal evolution of the carrier phase induced by satellite motion. As shown in Fig.~\ref{fig:cond_analysis}, LEO constellations exhibit significantly lower condition numbers compared to GNSS, indicating substantially improved conditioning. More importantly, the condition number for LEOs converges much faster than that for GNSS. Specifically, convergence is achieved within approximately 200 seconds for LEO satellites, whereas the condition number for GNSS does not converge even after 500 seconds of observation. Furthermore, it can be observed that, beyond a certain point in the observation window (approximately 120--130 seconds), the condition number for the LEO case begins to exhibit mild rippling behavior. This behavior arises from changes in the set of visible LEO satellites due to their low orbital altitude and high orbital velocity. As satellites enter and exit visibility, the UE begins acquiring carrier-phase measurements from newly visible satellites, thereby increasing the total number of integer ambiguities to be resolved. However, since the UE position has already converged to a relatively accurate estimate by this stage, the newly introduced ambiguities can be resolved rapidly, and the overall conditioning remains favorable. In contrast, the set of visible GNSS satellites remains largely unchanged over several minutes due to their much higher orbital altitude and slower orbital motion. Consequently, the number of integer ambiguities remains fixed over the observation window, and the conditioning improves only gradually, as temporal diversity increases at a significantly slower rate than in the LEO case.

It is worth noting that both the absolute condition numbers and the associated convergence times are relatively large in this analysis. This is because conditioning is evaluated under a carrier-phase-only positioning framework, where integer ambiguity resolution relies solely on temporal carrier-phase evolution and typically requires several minutes of observation. As noted in Section~\ref{sec:carrier_phase_tracking}, practical systems jointly exploit delay- or code-phase measurements to provide coarse range information and significantly reduce the integer ambiguity search space. The subsequent analysis demonstrates that incorporating delay measurements with carrier-phase measurements and exploiting faster satellite motion enables LEO-based systems to achieve ambiguity resolution and high-accuracy (sub-m) positioning within a few seconds.

\subsection{Integer Ambiguity Resolution}\label{sec:integer_ambgty_resol}
Next, we compare integer ambiguity resolution for GNSS- and LEO-based positioning using joint delay and carrier-phase measurements to enable rapid convergence. The total measurement duration is 3 seconds, with observation epochs spaced at 10~ms intervals. The primary objective of this study is to demonstrate how faster satellite motion in LEOs enables quicker convergence in carrier-phase positioning, especially when coupled with coarse delay measurements, and to assess the ability of LEO-based positioning to achieve cm-level accuracy. Accordingly, detailed carrier-phase measurement and tracking algorithms and their implementation-specific intricacies are not the main focus of this paper. Instead, we employ an error-modeling-based simulation framework, where delay and carrier-phase measurement errors are modeled using variances derived from the Cram\'er-Rao lower bound (CRLB). For the LEO case, Equivalent Isotropic Radiated Power (EIRP) density is configured to 34 dBW/MHz. The system operates in S-band at a 2 GHz carrier frequency (n256), with a 1 MHz bandwidth and a 15 kHz subcarrier spacing (SCS), consistent with NR-NTN specifications and validation methodologies~\cite{3gpp::38821}. For the GNSS case, we consider an operating frequency of 1575.42 MHz, GPS L1 coarse acquisition (C/A) or civilian code transmitted at 1.023 Mbps, and link budget numbers as given in~\cite{Edkgps2005}. For demonstration, we use a maximum of four participating satellites selected among the visible satellites based on the highest elevation angles, particularly relevant to LEO constellations designed primarily for communication rather than dedicated positioning.
\begin{figure}[!tb]
    \centering
    \includegraphics[width=0.88\linewidth]{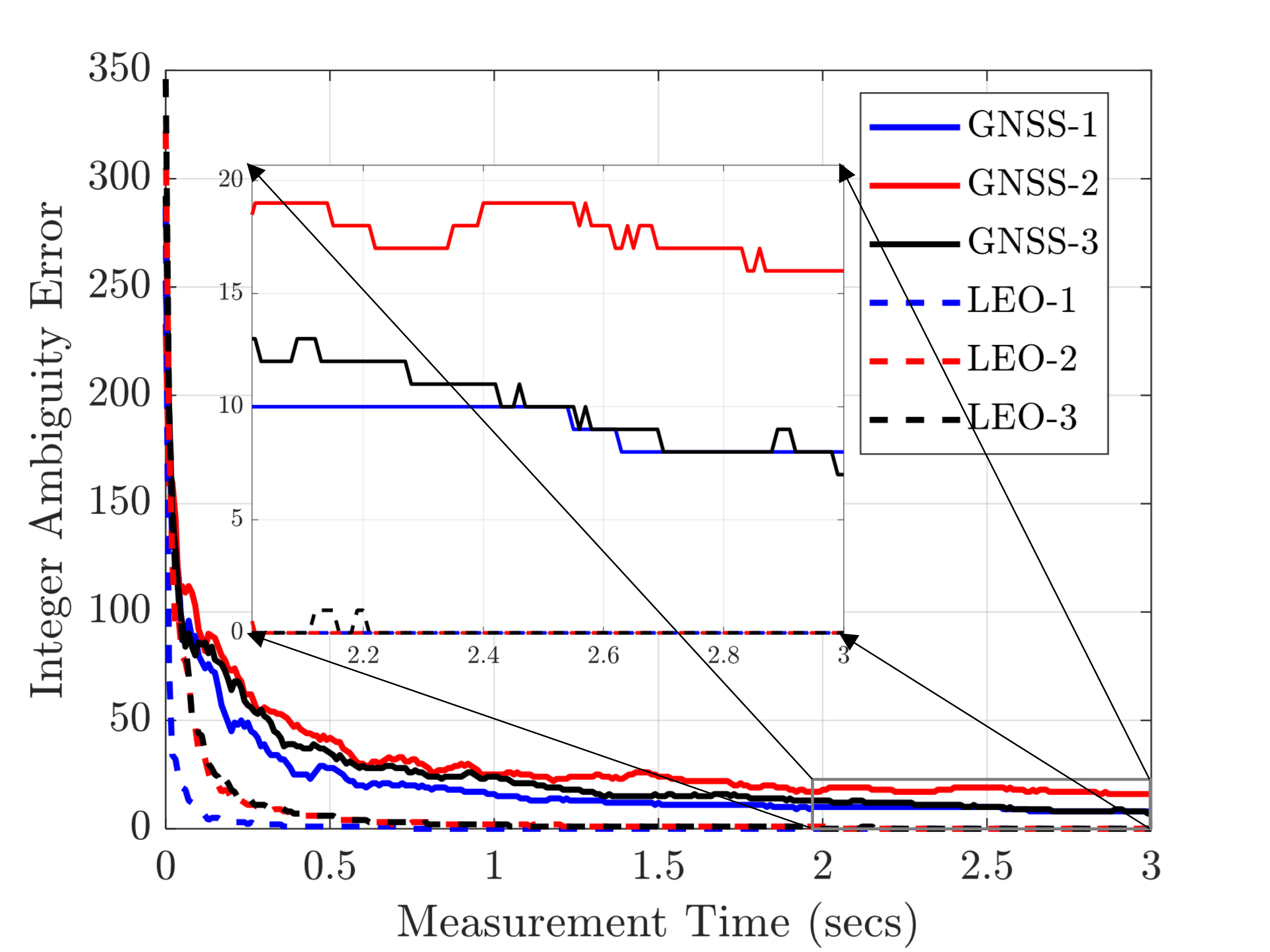}
    \captionsetup{font=small}
    \caption{{\em Integer Ambiguity Resolution: GNSS vs. LEO}: Integer ambiguity resolution over measurement time.}
    \label{fig:ambgty_resol}
\vspace{-16pt}
\end{figure}

Fig.~\ref{fig:ambgty_resol} depicts the integer ambiguity resolution performance as a function of measurement time for GNSS and LEO-based satellite positioning systems. Given that four satellites are considered and carrier-phase measurements are double-differenced, a total of three integer ambiguities must be resolved in both scenarios. As shown in the figure, joint exploitation of delay and carrier-phase measurements enables significantly faster ambiguity resolution in LEO constellations than in GNSS. In particular, for the LEO case, all integer ambiguities are resolved within 500 ms to 1 second, whereas for GNSS, ambiguity resolution is not achieved even with an observation window of 3 seconds. This is a significant result, as it clearly shows that faster satellite motion in LEOs provides greater temporal diversity in carrier-phase measurement acquisition and enables quicker convergence than GNSS. Additionally, the zoomed-in view in Fig.~\ref{fig:ambgty_resol} further reveals that, for the LEO case, all integer ambiguity errors converge to zero within a 3-second observation window, indicating successful and stable ambiguity resolution. In contrast, for GNSS, the ambiguity errors remain on the order of tens of carrier cycles over the same observation interval. The differences in ambiguity-convergence behavior between GNSS and LEO systems directly translate into differences in positioning accuracy and convergence time, which we discuss next.
\begin{figure}[t]
    \centering
    \resizebox{0.88\linewidth}{!}{% Resize to match text width
        \begin{minipage}{\linewidth} % Encapsulate subfigures in a minipage
            \centering
             \begin{subfigure}[b]{\linewidth}
                 \centering
                 \small
                 \includegraphics[width=\linewidth]{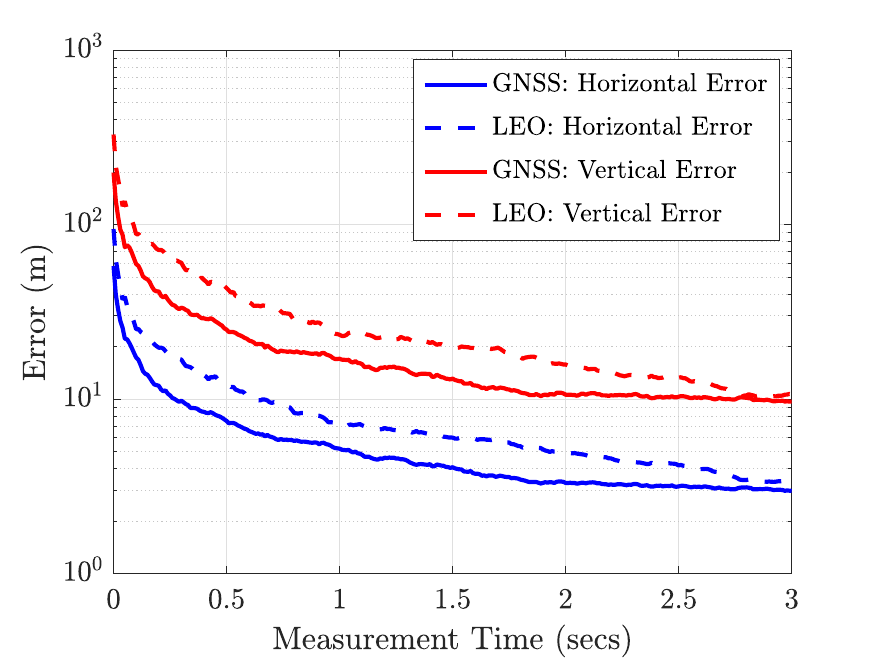}
                 \captionsetup{font=normal}
                 \caption{Delay-only positioning.}
                 \label{fig:delay_only_positioning}
             \end{subfigure}
            \hfill
             \begin{subfigure}[b]{\linewidth}
                 \centering
                 \small
                \includegraphics[width=\linewidth]{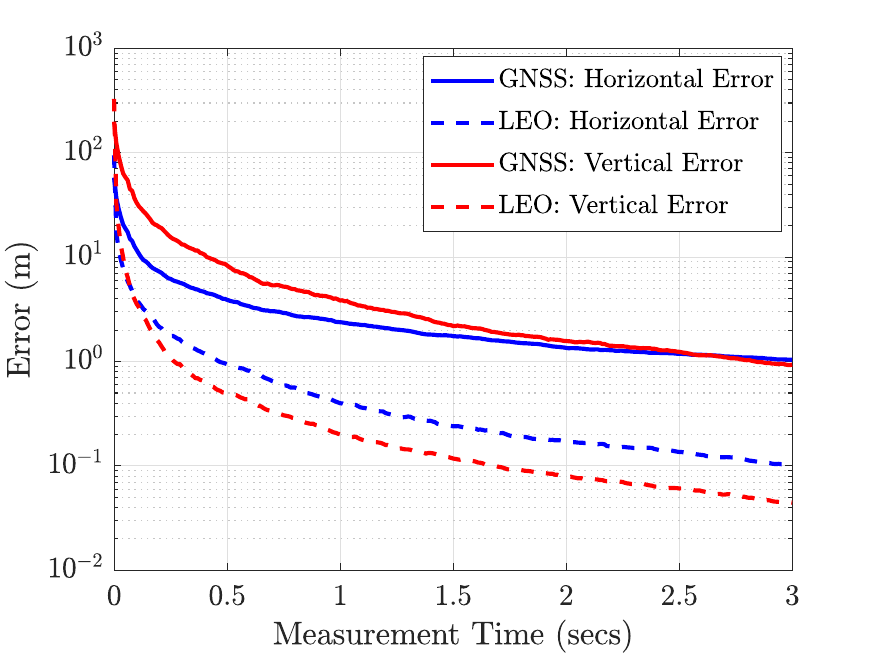}
                \captionsetup{font=normal}
                \caption{Joint delay-and-carrier-phase positioning.}
                \label{fig:joint_delay_carrier_phase_positioning}
             \end{subfigure}        
        \end{minipage}
    }
     \captionsetup{font=small}
     \caption{{\em Positioning performance:} Positioning performance comparison between LEO and GNSS considering delay-only and joint-delay-and-carrier-phase measurements.}
     \label{fig:positioning_performance}
     \vspace{-18pt} % Adjust the negative space as needed
\end{figure}

\subsection{Positioning Performance: Joint Delay-and-Carrier-Phase}\label{sec:positioning_performance}
Finally, we evaluate the positioning performance of GNSS- and LEO-based positioning systems. For this assessment, we reuse the same evaluation setup employed in the integer ambiguity resolution analysis. Two positioning scenarios are considered: 1) positioning using delay-only measurements and 2) positioning using joint delay-and-carrier-phase measurements. In both scenarios, measurements collected across multiple observation epochs are jointly processed to estimate the UE location. Recursive weighted least squares (RWLS)~\cite{1130282269234990848} is adopted as the positioning engine to enable efficient joint processing of delay and carrier-phase measurements acquired over multiple time epochs. \chb{For both LEO and GNSS, satellite ephemeris and clock information are assumed to be perfectly known or corrected, thereby allowing the comparison to isolate the impact of satellite dynamics and temporal geometric diversity on carrier-phase convergence. In practice, residual ephemeris and satellite clock errors can degrade carrier-phase accuracy and require appropriate estimation or correction, particularly for communication-centric LEO constellations.}

Fig.~\ref{fig:delay_only_positioning} presents the horizontal and vertical positioning errors obtained using delay-only measurements, while Fig.~\ref{fig:joint_delay_carrier_phase_positioning} illustrates the corresponding positioning performance achieved using joint delay-and-carrier-phase measurements for both LEO and GNSS constellations. As observed, GNSS performs better than LEO in the delay-only case, primarily because of its favorable satellite geometry, which is particularly optimized for dedicated positioning services. However, neither system achieves cm-level accuracy when relying solely on delay measurements. It is important to note that this analysis considers a single PRS transmission for delay estimation. In our prior work~\cite{dureppagari_leo_11049853}, we demonstrated that LEO-based positioning can achieve sub-10m accuracy using multi-symbol PRS transmissions, with best-case performance approaching the meter-level using delay-only measurements. In contrast, Fig.~\ref{fig:joint_delay_carrier_phase_positioning} demonstrates that by jointly exploiting delay and carrier-phase measurements, LEO-based positioning systems can achieve cm-level accuracy within a 500-ms observation window. This rapid convergence is consistent with the integer ambiguity resolution behavior shown in Fig.~\ref{fig:ambgty_resol}. Although GNSS positioning performance also improves when carrier-phase measurements are incorporated alongside delay measurements, ambiguity errors remain on the order of tens of carrier cycles even after a 3-second observation window. These residual ambiguity errors result in range estimation errors on the order of meters, thereby limiting the achievable positioning accuracy. In addition to convergence analysis, these results further validate that the faster satellite motion inherent to LEOs not only enables quicker convergence but also helps achieve cm-level accuracy within a short observation window by providing higher temporal diversity in carrier-phase acquisition than GNSS. 

\chb{We consider a maximum of four satellites to maintain a consistent evaluation setup across LEO and GNSS and to assess carrier-phase positioning with a limited number of satellites for positioning, as may occur in communication-centric LEO-NTN deployments. While additional satellites can provide greater measurement and geometric diversity, our evaluations exhibit similar relative convergence trends with more satellites considered for positioning and do not alter the key findings presented in this article.} %\chb{From a practical deployment perspective, the proposed framework assumes the availability of sufficiently accurate satellite ephemeris and clock information, phase-coherent carrier tracking across observation epochs, and appropriate cycle-slip detection and recovery. The continuous phase-tracking waveform can be accommodated using only one PRB, either within suitable guard/edge resources or the active NR-NTN bandwidth, and can be supported through a dedicated phase-stable RF chain or integrated with the existing satellite transmit chain. Its bandwidth and EIRP should be configured to maintain reliable phase and Doppler tracking while limiting the impact on communication resources.}
% \begin{figure}[t]
%     \centering
%     \resizebox{\linewidth}{!}{% Resize to match text width
%         \begin{minipage}{\linewidth} % Encapsulate subfigures in a minipage
%             \centering
%              \begin{subfigure}[b]{0.9\linewidth}
%                  \centering
%                  \small
%                  \includegraphics[width=\linewidth]{figures/results/gnss_vs_leo_delay_only_1.pdf}
%                  \captionsetup{font=small}
%                  \caption{Delay-only positioning.}
%                  \label{fig:delay_only_positioning}
%              \end{subfigure}
%             \hfill
%              \begin{subfigure}[b]{0.9\linewidth}
%                  \centering
%                  \small
%                 \includegraphics[width=\linewidth]{figures/results/gnss_vs_leo_delay_phase_1.pdf}
%                 \captionsetup{font=small}
%                 \caption{Joint delay-and-carrier-phase positioning.}
%                 \label{fig:joint_delay_carrier_phase_positioning}
%              \end{subfigure}        
%         \end{minipage}
%     }
%      \captionsetup{font=small}
%      \caption{{\em Positioning performance:} Positioning performance comparison between LEO and GNSS considering delay-only and joint-delay-and-carrier-phase measurements.}
%      \label{fig:positioning_performance}
%      \vspace{-16pt} % Adjust the negative space as needed
% \end{figure}

\section{Research Landscape and Open Problems}\label{sec:conclusions}
While the results presented in this paper are encouraging, several fundamental research challenges and system-level aspects must be addressed before carrier-phase-enabled LEO positioning can be realized in practical NR-NTN deployments.

{\em Phase Coherence and Cycle-Slip Mitigation.} Reliable carrier-phase positioning requires sufficient phase coherence of the transmitted waveform. However, communication-focused LEO constellations are not designed for GNSS-grade phase stability, and oscillator impairments, payload switching, beam hopping, low SNR, and multipath can introduce phase discontinuities and cycle slips. Future work should quantify the required phase-coherence and phase-noise limits and develop robust cycle-slip detection and recovery techniques for maintaining ambiguity continuity under practical LEO-NTN conditions.

{\em Integer Ambiguity Resolution.} Although rapid satellite motion in LEOs accelerates ambiguity convergence, reliable integer ambiguity resolution remains challenging due to intermittent PRS transmissions and high Doppler rates. Future work should focus on robust ambiguity fixing and validation methods that exploit joint delay-and-carrier-phase information under realistic NR-NTN constraints.

{\em Extension to Kinematic and Network-Based Positioning.} This study focused on a conventional static positioning framework to isolate convergence behavior. \chb{For kinematic UEs, UE-induced Doppler and phase variations must be jointly tracked with satellite-induced dynamics while maintaining phase continuity and reliable integer ambiguity resolution. Dynamic state-space estimators that jointly track UE motion, carrier phase, Doppler, and ambiguity states, potentially aided by the proposed dual-waveform design, present a promising direction for future research.} \chb{Network-based positioning can also leverage future free-space optical (FSO)-based inter-satellite links to exchange precise timing, ephemeris, and clock-correction information across the LEO constellation, thereby supporting synchronization and high-accuracy carrier-phase positioning over the NR-NTN radio interface.}

% {\em Multi-LEO and Hybrid GNSS--LEO Carrier-Phase Fusion.} While this work considered repurposed communication-focused constellations, future systems are likely to involve multi-LEO cooperation and hybrid GNSS--LEO carrier-phase fusion, particularly in challenging environments with partial visibility. Designing ambiguity-consistent fusion frameworks across heterogeneous constellations and orbital regimes is a promising yet nontrivial research direction.

{\em Standardization and NR-NTN Integration.} From a standards perspective, enabling carrier-phase positioning raises important questions regarding PRS design, signaling support, waveform multiplexing, resource and power allocation aspects, and receiver requirements in NR-NTN. The dual-waveform concept introduced in this work provides a potential starting point, but its standardization implications require further investigation.

\bibliographystyle{IEEEtran}
\bibliography{hokie}
\end{document}

%% file: notation.tex
\def\nb0{{\mathbf{0}}}
\def\nb1{{\mathbf{1}}}